\documentclass[aps,prb,twocolumn,groupedaddress,showpacs,showkeys]{revtex4-1}

\usepackage{graphicx}
\usepackage{amsmath}
\usepackage{amssymb}

\begin{document}

\title{Motor effect in electron transport through a molecular junction with torsional vibrations}

\author{Ivan A. Pshenichnyuk}
\author{Martin \v{C}\'{\i}\v{z}ek}
\affiliation{Institute of Theoretical Physics, Faculty of Mathematics and Physics, Charles University, Prague, Czech Republic}
\date{\today}

\begin{abstract}
 We propose a model for a molecular junction with internal anharmonic torsional vibrations interacting
 with an electric current. The Wangsness-Bloch-Redfield master equation approach is used to determine the 
 stationary reduced density matrix of the molecule. The dependence of the current, excitation energy
 and angular momentum of the junction on the applied voltage is studied.
 Negative differential conductance is observed in the current-voltage characteristics. It is 
 shown that a model with vibrationally dependent coupling to the electrodes, asymmetric with respect
 to the interchanging of electrodes, leads to a strong correlation between the applied voltage and the angular
 momentum of the junction. The model thus works as a molecular motor, with the angular momentum controlled 
 by the size and sign of the voltage. 
\end{abstract}


\maketitle

\section{Introduction}

Molecular electronics has become a dynamically growing field in the last decade. 
It is a highly interdisciplinary research topic presenting many challenges in both theoretical and synthetic chemistry, solid state and many particle physics and non-equilibrium statistical physics.
Much attention has been paid to the careful preparation of single molecule junctions and to studies of the conductive properties of different molecules.
Numerous papers \cite{selzer-2006,galperin-2008,galperin-2007,lu-2010}, studying both the experimental and theoretical aspects, have already been devoted to the subject. It has also been noted \cite{jorn-2010, benesch-2009, venkataraman-2006, mishchenko-2010} 
that the molecular structure and vibrations or conformational changes are among the main points of interest, making the molecules distinct from other electronic elements.

The coupling of electronic and mechanical degrees of freedom is a standard part of electrical engineering and the resulting gadgets are a standard 
part of our life. In recent decades the coupling of electronic and vibrational degrees of freedom has been achieved in nanoscale 
solid state devices (see \citet{schwab-2005} and \citet{blencowe-2004} for reviews on NEMS) with the quantum regime reached in the mechanical
degree of freedom. Driving the molecular vibrations with the electronic current is a natural extension of this concept.  
We can consider a molecular electronic element or molecular junction with some molecular groups performing rotational motion in response to the bias voltage across the molecular junction. Such elements have already been anticipated \cite{wang-2008} and the work is closely related to electron shuttles \cite{novotny-2003, gorelik-1998}. Recently, the excitation of periodic nuclear motion in molecular
junctions has been studied as a classical motion of atoms in non-conservative forces induced by the current flow \cite{dundas-2009, lu-2010}.
Artificially-built molecular motors, where molecular vibrations are driven by light or stochastic fluctuations due to interaction with a thermal bath (Brownian motion), have also been studied. The externally-driven torsional motion of some small parts of molecules has been demonstrated for molecules both in gas \cite{madsen-2009} and mounted on surfaces \cite{horansky-2007}.

The main goal of this paper is to demonstrate that the rotational motion of a molecular group in a metal-molecule-metal junction should in fact be a very common phenomenon and that there are only two conditions required: 1) the presence of some part of the molecule capable of rotation with a moderately small potential barrier against this rotation; and 2) a breaking of the mirror symmetry in the junction (chirality of the junction).

To achieve this goal we set up a general model for the description of the interaction of a current flowing through a molecule with anharmonic molecular vibrations. To analyze and understand the effect of anharmonicity in more-or-less well-controlled conditions we first define a model with some small vibrational coupling due to a small vibrational potential energy shift. We then switch to a more realistic model of molecular vibrations motivated by real molecular rotors as used in a previous experiment with light-driven artificial molecular motors \cite{horansky-2007}. The dynamics of the system will be studied using rate equations for the reduced density matrix of the molecule. The current through the junction and the average angular momentum of the molecule are calculated as a function of the voltage drop across the junction.

\section{Model}

The molecular junction consists of two metallic electrodes or leads L and a molecular bridge M connected between them.
The corresponding division of the Hamiltonian reads
\begin{equation}
\label{HamiltonianPartitioning}
  H = H_M+H_L+H_{ML}.
\end{equation}
The model of the bridging molecule is assumed to consist of one vacant electronic level that allows electrons
to tunnel through it from the left lead to the right lead.
We will call this state the localized state. There is an extra charge on the molecule when
this state is occupied, and we will thus speak about a neutral, or a charged, molecule (anion) if the state
is unoccupied or occupied, respectively.
In addition, we include a vibrational degree of freedom on the bridge, which exchange energy with the electrons.
We thus assume
\begin{equation}
  H_M = h_0 dd^{\dagger} + h_1 d^{\dagger}d,
\end{equation}
where $d^{\dagger}$ and $d$ create and annihilate electrons in the localized state on the bridge. The operators
$h_0$ and $h_1$ describe the vibrational motion of the nuclei in the neutral and charged state respectively.
The form of both $h_0$ and $h_1$ is based on the Born-Oppenheimer vibrations for isolated molecules and in the spirit
of the Born-Oppenheimer approximation we assume, that both $h_0$ and $h_1$ commute with $d^{\dagger}$ and $d$.
The Born-Oppenheimer approximation is, of course, broken when we allow electronic transitions due to the
coupling to the leads $H_{ML}$.

\begin{figure}
\centerline{\includegraphics[width=3.00in]{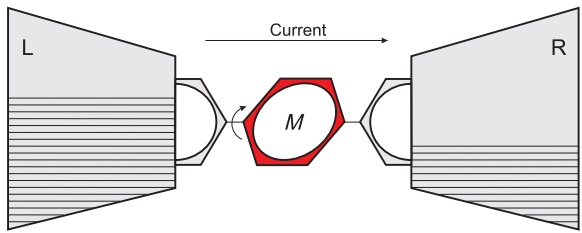}}
\caption{Schematic representation of the junction\label{junction-schem}}
\end{figure}
Figure~\ref{junction-schem}, showing schematically the system that we have in mind, may serve to guide us through
further specifications of the model. We will consider only one nuclear (vibrational) degree of freedom
representing the rotation of a part of the molecule, pictured as a benzene ring bound in para positions
to other parts of the molecule that are directly attached to the metal leads.
The vibrational Hamiltonians $h_0$ and $h_1$ are both of the form
\begin{equation}
  h_i= - \frac{1}{2I} \frac{\partial^2}{\partial\varphi^2} + V_i(\varphi), \\
\end{equation}
where $\varphi\in\langle 0,2\pi\rangle$ represents the vibrational coordinate (angle of rotation of the ring), $I$ is
the moment of inertia of the ring and $V_i(\varphi)$ is the Born-Oppenheimer vibrational potential
of the molecule with unoccupied ($i=0$) or occupied ($i=1$) electronic level. In this paper
we want to characterize the main features of the dynamics of transport through
junctions with such anharmonic vibrations and we assume a simple analytic yet rather
general shape of potentials \cite{petreska-2008} as
\begin{equation}
  V_i(\varphi) = \varepsilon_i + A_i \cos (n_i\varphi + \varphi_i).
\end{equation}
We use two sets of models. In the models of group 1 we want to capture the basic features
of the usual models used for studing molecular conduction junctions coupled to vibrations
\cite{glazman-1988, braig-2003, cizek-2004, galperin-2006}, but
to allow for large amplitude anharmonic motion. We thus take both $V_0$ and $V_1$ to be
the potentials of the mathematical pendulum ($n_0=n_1=1$), one shifted with respect to the other
($\varepsilon_0$, $\varphi_0$ differs from $\varepsilon_1$, $\varphi_1$). For these
models we also set the moment of inertia $I$ to an unrealistically small value
\footnote{The value is too small for the phenyl-based rotor, but is comparable, for example, to the moment of inertia of the methyl group.}
to allow for a more efficient numerical solution.
The second group of models are motivated by more
realistic parameters, as might be expected in real molecular systems
(see, for example, \citet{fortrie-2007} or \citet{tsuzuki-1999}). The vibrational potential
$V_0$ for the neutral molecule is thus characterized with a smaller amplitude $A_0$
and larger number of oscillations ($n_0=2$) than the potential for the charged
molecule $V_1$ (for biphenyl junction see \citet{cizek-2005}). Also the value of $I$ is set larger to describe
the inertia of the benzene ring. All values of the model parameters are summarized in Table \ref{models-parameters}.

In principle, the parameters of the molecular Hamiltonian should also depend on the voltage
applied across the junction. We follow the common practice (see, for example, \citet{galperin-2006} or \citet{hartle-2009}
and the works cited there) of using the molecular Hamiltonian independent of voltage.
This approximation is also supported by the calculation of molecular potentials 
in the electric field made by \citet{petreska-2008}, where the potential barriers do not 
show a significant change for realistic values of the electric field.

\begin{table*}
  \caption{Summary of all parameters for the models in use. The energies $\varepsilon_i$, $A_i$ are in units of eV, angles in radians and moment of inertia in atomic units ($m_e a_0^2$).\label{models-parameters}}
  \begin{ruledtabular}
  \begin{tabular}{cccccccccc}
          & & \multicolumn{3}{c}{$V_0(\varphi)=\varepsilon_0+A_0\cos(n_0\varphi)$}
            & \multicolumn{3}{c}{$V_1(\varphi)=\varepsilon_1+A_1\cos(\varphi+\varphi_1)$}
            & \multicolumn{2}{c}{$V_{\alpha}(\varphi)=\cos(\varphi-\varphi_{\alpha})$} \\
    \colrule
 Model & $I$    & $\varepsilon_0$ & $A_0$ & $n_0$ & $\varepsilon_1$ & $A_1$ & $\varphi_1$ & $\varphi_l$ & $\varphi_r$ \\
    \colrule
    1a &        &                 &       &       &                 &       &             &  -          &  -      \\
    1b & 2000   & 1.25            & 1.25  & 1     & 1.35            & 1.25  & 0.03        & $\pi$       & $\pi$   \\
    1c &        &                 &       &       &                 &       &             & $\pi$       & 3$\pi$/2 \\
    \colrule
    $\begin{array}{c} 2a \\ 2b \end{array}$
       & 226852 & -0.05           & 0.05  & 2     & 0.10            & 0.20  & $\pi$+0.03  & 0           &
       $\begin{array}{c} 1.0 \\ 2.0 \end{array}$ \\
  \end{tabular}
  \end{ruledtabular}
\end{table*}

The Hamiltonian of the leads is written in the form
\begin{equation}\label{e:leadH}
  H_L = \sum_{\alpha=l,r} \sum_{k}  \varepsilon_{k\alpha}c_{k\alpha}^{\dagger}c_{k\alpha},
\end{equation}
where the operator $c_{k\alpha}^{\dagger}$ creates an electron in the state with a wave number $k$ in the lead
$\alpha\in\{l,r\}$. To be more specific we assume that the Hamiltonian (\ref{e:leadH}) was obtained
from the diagonalization of the one-dimensional nearest-neighbor tight-binding model \cite{ashcroft}, i.\ e.\
the dispersion relation reads
\begin{equation}
  \label{e:TBMSpectrum}
  \varepsilon_{k\alpha}=\mu_{\alpha} + 2\beta_1 \cos(k),
\end{equation}
where $\mu_{\alpha}$ is the chemical potential of the lead $\alpha$. When we talk about the voltage $U$ applied to the junction
we assume $\mu_l=+\frac{U}{2}$ for the left lead and $\mu_r=-\frac{U}{2}$ for the right lead. The parameter $\beta_1 = 3\,\text{eV}$
defines the width of the conduction band in the leads. It is chosen in order to provide a band, wide enough to exclude edge effects.
Each lead is separately assumed to be in thermodynamic equilibrium with states populated according to the Fermi-Dirac distribution
\begin{equation}
\label{e:TBMPopulations}
f_{\alpha}(\varepsilon_{k\alpha}) = \left[1+e^{(\varepsilon_{k\alpha}-\mu_{\alpha})/kT}\right]^{-1}.
\end{equation}

To complete the description of the model the molecule-lead coupling 
\begin{equation}
  H_{ML} = \sum_{\alpha=l,r} \sum_k V_{dk\alpha}(d^{\dagger}c_{k\alpha}+c_{k\alpha}^{\dagger}d)
\end{equation}
must be specified.
We assume the separable form of the coupling coefficients $V_{dk\alpha}=V_{k}V_{\alpha}(\varphi)$.
The dependence on the electron momentum $V_k=\beta_2\sin(k)$ is again motivated by the one dimensional
nearest-neighbor tight binding model, where the sine term comes from the electronic wavefunction \cite{cizek-2004}
and the parameter $\beta_2$ is the overall coupling strength, which is set to $\beta_2 = 0.07 \, \text{eV}$.
We shall study several forms of the angle dependent part $V_{\alpha}(\varphi)$.
\begin{itemize}
\item Model 1a. The case of {\em angle-independent} coupling $V_{\alpha}(\varphi) = 1$. This assumption simplifies the treatment
    of the dynamics significantly. Furthermore, this is the case most often considered by other studies 
    \cite{glazman-1988, braig-2003, cizek-2004, galperin-2006}.
\item Model 1b. Here we introduce {\em angle-dependence} into the coupling term $V_{\alpha}(\varphi) = \cos(\varphi  -\varphi_{\alpha})$
    but we do not break the symmetry between the left and right lead $\varphi_{l} = \varphi_{r} = \pi$. The form of the angular
    dependence can be motivated by the H\"{u}ckel model (see \citet{cizek-2005,pauly-2008}).
\item Model 1c. Finally we {\em break the symmetry} of the system, taking $V_{\alpha}(\varphi)$ in the same form as in model 1b but with $\varphi_{l} = \pi$, $\varphi_{r} = 3\pi/2$. The asymmetry can arise as a result of different molecular bonding
    to the left and right lead. We can, for example, consider a molecule consisting of a chain of three aromatic rings with
    the first ring fixed to the left lead the middle ring acting as the rotor, and the right ring fixed to the right lead (this
    idea is used in the diagram of the model in Figure \ref{junction-schem}).
    Breaking the symmetry provides circumstances for observing the "motor effect", i.e. preferential rotation of the
    rotor in one direction depending on the sign of the voltage applied across the junction.
\item Models 2a and 2b use the same form of coupling as the Model 1c. The parameters $\varphi_{l}$, $\varphi_{r}$
    for all models are summarized in Table \ref{models-parameters}.
\end{itemize}

\section{Theory}

\begin{figure}
\centerline{\includegraphics[width=3.00in]{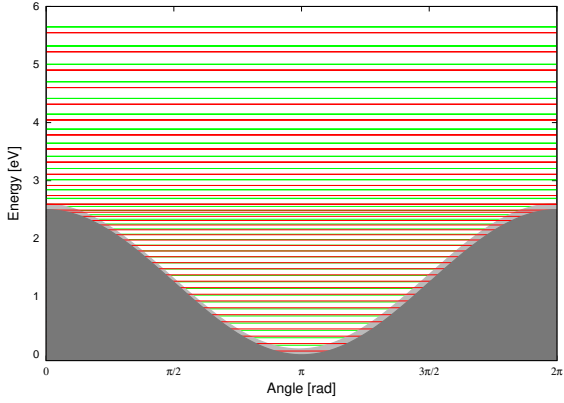} }
\caption{Energy levels of the unoccupied (red lines) and occupied (green lines) molecular bridge of Model 1. The respective potentials are the borders of dark gray and light gray areas.\label{m1_potentials}}
\end{figure}

\begin{figure}
\centerline{\includegraphics[width=2.90in]{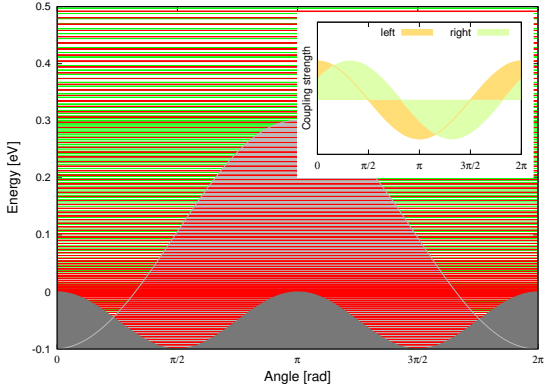} }
\caption{The energy levels and potentials for Model 2 plotted in the same way as in the previous figure. The angular dependence of molecule-lead couplings $V_{\alpha}(\varphi)=\cos(\varphi-\varphi_{\alpha})$ are shown in the inset for the left $\alpha=l$ and right $\alpha=r$ leads. \label{m2_potentials}}
\end{figure}

The Hilbert space of our system is a direct product of the spaces of electronic and vibrational degrees of freedom.
In the electronic space, we can define a complete set of projectors
\begin{equation}
   dd^{\dagger} + d^{\dagger}d = 1,
\end{equation}
where the projector $dd^{\dagger}$ projects on the part of the Hilbert space with the unoccupied molecular bridge
and $d^{\dagger}d$ projects on the occupied bridge space. It is advantageous to use different basis sets in the vibrational
part of the Hilbert space according to the occupation of the bridge. We thus define two basis sets $|n\rangle$
and $|v\rangle$
\begin{equation}
\label{e:h-s_equations}
\begin{array}{c}
  h_0|n\rangle = E_n|n\rangle , \\
  h_1|v\rangle = E_v|v\rangle.
\end{array}
\end{equation}
This choice diagonalizes the molecular part of the Liouvillian operator $\mathcal{L}_0$ (see next section) and
in the case of the harmonic molecular potentials $V_i$ it is equivalent to performing the polaron transform
\footnote{The polaron transform performs the spatial shift in the coordinate $\varphi$ to convert the basis states $|n{\rangle}$ into basis states $|v{\rangle}$. }
as made in the independent boson model \cite{mahan}.
With cosinusoidal potentials, the states $|n\rangle$, $|v\rangle$ can be expressed in terms of Mathieu functions. Energy levels $E_n$ and $E_v$ (n,v = 0,1,...) of the unoccupied and occupied molecule for models 1 and 2 are shown in Figures \ref{m1_potentials} and \ref{m2_potentials} respectively. From the point of view of classical mechanics, such a shape of the potential provides a rotational barrier with an energy of $\epsilon_i+A_i$, which divides the two types of motion: vibrational (when the energy of the system is below the barrier) and rotational. In a quantum mechanical description the wavefunctions of states with an energy below the barrier are localized in space and in this sense can be called vibrational. In contrast, states above the barrier are delocalized through the whole interval $\langle 0,2\pi \rangle$. These states are two times degenerate, as the two directions of rotation are available above the barrier and they are energetically equivalent. We will call them rotational states. Well above the classical barrier, when the system has a lot of energy and does not feel the potential anymore, the states almost coincide with the free rotor states.

\subsection{Master Equation}

Different theoretical approaches to the transport of charge across a molecular junction with a coupling to vibrations have been discussed in the review article of \citet{galperin-2007}. Here we consider a weak-coupling case, which can be treated with the standard master-equation approach \cite{lehmann-2004, mitra-2004, hartle-2009}. A slight modification is needed to account for the anharmonicity of the vibrations as described in what follows.

In the framework of master equation (ME) theory we calculate the reduced density matrix (RDM) of the molecular bridge $\rho$.
We start from ME in the Wangsness-Bloch-Redfield (WBR) form \cite{timm-2008}
\begin{equation}
\label{WBR-ME}
\frac{\partial}{{\partial}t} \rho(t) = \mathcal{L}_0[\rho(t)] + \mathcal{L}_1[\rho(t)],
\end{equation}
where the Liouvillian super operator $\mathcal{L} \equiv \mathcal{L}_0 + \mathcal{L}_1$ with
\begin{equation}
\label{eq:Liouv}
\begin{array}{c}
\mathcal{L}_0[\rho(t)] = -i[H_M,\rho(t)] , \\
\mathcal{L}_1[\rho(t)] = -\text{Tr}_L \int\limits_0^{\infty} d\tau
[H_{ML},[H_{ML}(-\tau), \rho(t) \otimes \rho_L^0]] ,
\end{array}
\end{equation}
and
\begin{equation}
H_{ML}(-\tau) = e^{-i (H_M + H_L)\tau} H_{ML} e^{+i (H_M + H_L)\tau}.
\end{equation}
These equations are derived as a second order expansion in $H_{ML}$, which is assumed to be small.
The equilibrium RDM of the leads is denoted $\rho_L^0$. The first term $\mathcal{L}_0$ defines the time evolution for a "free" molecular bridge that is disconnected from the leads. The second term $\mathcal{L}_1$ incorporates the influence of the leads and could be written as the sum of two independent terms $\mathcal{L}_1=\mathcal{L}_{1,l}+\mathcal{L}_{1,r}$ for the left and right leads respectively.

Here we are not interested in the time evolution of the density matrix. We assume the existence of a stationary state $\partial\rho/\partial t =0$
which satisfies the equation $\mathcal{L}[\rho] = 0$. In basis representation, the Liouvillian is a 4th rank tensor. We search for a nontrivial solution of the equation
\begin{equation}
\label{ME-tensor4}
\sum\limits_{kl} \mathcal{L}_{ijkl} \rho_{kl} = 0.
\end{equation}

Before we write expressions for the components of this tensor we decompose RDM $\rho$ in the following way
\begin{equation}
\label{RDMdecomposition}
\rho = \rho_{00}dd^{\dagger} + \rho_{11}d^{\dagger}d
     + \rho_{01}d + \rho_{10}d^{\dagger}.
\end{equation}
With this representation the right hand side of the equation (\ref{WBR-ME})
can be reorganized in blocks
\begin{equation}
\label{ME-mod}
\begin{array}{c}
\mathcal{L} \begin{pmatrix} \rho_{00} \\ \rho_{11} \end{pmatrix} =
\left[
\begin{pmatrix} \mathcal{L}^0_{00} & 0 \\ 0 & \mathcal{L}^0_{11} \end{pmatrix} +
\begin{pmatrix} \mathcal{L}^{l}_{00} & \mathcal{L}^{l}_{01} \\
                \mathcal{L}^{l}_{10} & \mathcal{L}^{l}_{11} \end{pmatrix}  \right. \\
\left. +
\begin{pmatrix} \mathcal{L}^{r}_{00} & \mathcal{L}^{r}_{01} \\
                \mathcal{L}^{r}_{10} & \mathcal{L}^{r}_{11} \end{pmatrix}
\right]
\begin{pmatrix} \rho_{00} \\ \rho_{11} \end{pmatrix}.
\end{array}
\end{equation}
Equations for $\rho_{01}$ and $\rho_{10}$ are decoupled from this system for $\rho_{00}$ and $\rho_{11}$
and they are not included here, since the observables of interest are also independent of $\rho_{01}$ and $\rho_{10}$
(see below).
To write the elements explicitly we use basis sets (\ref{e:h-s_equations}). Basis
$\{|n\rangle\}$ for the block $\rho_{nn'} \equiv \langle n|\rho_{00}|n'\rangle$
and $\{|v\rangle\}$ for the block $\rho_{vv'} \equiv \langle v|\rho_{11}|v'\rangle$.
We will consistently use the letters $n$, $n_1$, $n_2$, $n'$ to number the vibrational states
of the unoccupied molecule and the letters $v$, $v_1$, $v_2$, $v'$ for the states of the occupied
molecule, omitting the index 0/1 that would otherwise distinguish the occupancy.
Thus the components
$\mathcal{L}^{0}_{n'_1 n'_2 n_1 n_2}$, $\mathcal{L}^{0}_{v'_1 v'_2 v_1 v_2}$
of the zeroth order contributions $\mathcal{L}^0_{00}$, $\mathcal{L}^0_{11}$ read
\begin{equation}
\label{L0-tensor}
\begin{array}{c}
\mathcal{L}^0_{n'_1,n'_2,n_1,n_2} = i\delta_{n'_2n_2}\delta_{n'_1n_1}(E_{n_2}-E_{n_1}) ,  \\
\mathcal{L}^0_{v'_1,v'_2,v_1,v_2} = i\delta_{v'_2v_2}\delta_{v'_1v_1}(E_{v_2}-E_{v_1}) .
\end{array}
\end{equation}
The lowest order contributions $\mathcal{L}^{\alpha}_{ij}$ describing the presence
of the leads $\alpha=l,r$ in eq.~(\ref{ME-mod}) are expressed as
\begin{widetext}
\begin{equation}
\label{L1-tensor}
\begin{array}{l}
\mathcal{L}^{\alpha}_{n'_1,n'_2,n_1,n_2} =
   - \frac{1}{2}\delta_{n_1n'_1} \sum\limits_v f_{\alpha}(\omega_{n_2 v}) \Gamma_{\alpha}(\omega_{n_2 v})V^{\alpha}_{n_2 v} V^{\alpha}_{v n'_2}
   - \frac{1}{2}\delta_{n_2n'_2} \sum\limits_v f_{\alpha}(\omega_{n_1 v}) \Gamma_{\alpha}(\omega_{n_1 v})V^{\alpha}_{v n_1} V^{\alpha}_{n'_1 v} ,\\
\mathcal{L}^{\alpha}_{v'_1,v'_2,v_1,v_2} =
   - \frac{1}{2}\delta_{v_1v'_1} \sum\limits_n [1-f_{\alpha}(\omega_{n v_2})] \Gamma_{\alpha}(\omega_{n v_2})V^{\alpha}_{v_2 n} V^{\alpha}_{n v'_2}
   - \frac{1}{2}\delta_{v_2v'_2} \sum\limits_n [1-f_{\alpha}(\omega_{n v_1})] \Gamma_{\alpha}(\omega_{n v_1})V^{\alpha}_{n v_1} V^{\alpha}_{v'_1 n} ,\\
\mathcal{L}^{\alpha}_{n'_1,n'_2,v_1,v_2} =
     \frac{1}{2}[1-f_{\alpha}(\omega_{n'_1 v_1})] \Gamma_{\alpha}(\omega_{n'_1 v_1})V^{\alpha}_{n'_1 v_1} V^{\alpha}_{v_2 n'_2}
   + \frac{1}{2}[1-f_{\alpha}(\omega_{n'_2 v_2})] \Gamma_{\alpha}(\omega_{n'_2 v_2})V^{\alpha}_{v_2 n'_2} V^{\alpha}_{n'_1 v_1} ,\\
\mathcal{L}^{\alpha}_{v'_1,v'_2,n_1,n_2} =
     \frac{1}{2}f_{\alpha}(\omega_{n_1 v'_1}) \Gamma_{\alpha}(\omega_{n_1 v'_1})V^{\alpha}_{v'_1 n_1} V^{\alpha}_{n_2 v'_2}
   + \frac{1}{2}f_{\alpha}(\omega_{n_2 v'_2}) \Gamma_{\alpha}(\omega_{n_2 v'_2})V^{\alpha}_{n_2 v'_2} V^{\alpha}_{v'_1 n_1},
\end{array}
\end{equation}
\end{widetext}
where $\omega_{nv}\equiv E_v-E_n$ is the transition energy, $V^{\alpha}_{nv} \equiv \langle n|V_{\alpha}(\varphi)|v \rangle$
is the generalized Franck-Condon overlap for the transition and the factors
$f_{\alpha}(\omega) \Gamma_{\alpha}(\omega)$ and $[1-f_{\alpha}(\omega)] \Gamma_{\alpha}(\omega)$
come from the imaginary part of the functions
\begin{equation}
\label{SE-definition}
\begin{array}{c}
\xi_1 (E) \equiv
i \int\limits_0^{\infty}d\tau \sum\limits_{k} e^{-i({\varepsilon_{k\alpha}}-E)\tau} (f_{k\alpha}) V_k^2 ,\\
\xi_2 (E) \equiv
i \int\limits_0^{\infty}d\tau \sum\limits_{k} e^{+i({\varepsilon_{k\alpha}}-E)\tau} (1-f_{k\alpha}) V_k^2
\end{array}
\end{equation}
resulting from the time integration in eq.~(\ref{eq:Liouv}).
The real part of these functions that leads to renormalization of the energy levels is neglected
here (see also discussion in \citet{leijnse-2008} where it is argued that the real part is canceled out in higher orders). The imaginary parts can be calculated analytically. For the tight binding model of leads $\Gamma(E)$ become \cite{cizek-2004}
\begin{equation}
\label{SE-Gamma}
\Gamma(E)= \frac{\beta_2^2}{\beta_1^2}
\sqrt{4\beta_1^2 - (E-\mu_{\alpha})^2}
\end{equation}
inside the band (i.e. when $E \in [\mu - 2\beta_1,\mu+2\beta_1]$) and equal to zero outside.

It is often argued that the nondiagonal elements ($\rho_{nn'}$ and $\rho_{vv'}$ for $n\neq n'$, $v\neq v'$ ) in RDM
decay rapidly in time and are neglected in the search for the stationary state. Here the nondiagonal elements
for near-degenerate states lead to a nonzero angular momentum for the ring and can not be neglected.
However, to increase the numerical efficiency we consider only the elements close to the diagonal and assume that the RDM
has a band structure. Final results are presented for the RDM, which includes 14 sub diagonals, while 
its overall dimensions are 202$\times$202 for Model 1 and 402$\times$402 for Model 2. 
We tested that the presented results are stable with respect to changes in the number
of sub diagonals and the number of basis functions used in the calculation.

For numerical convenience, we reshape the tensors $\mathcal{L}$ of the fourth rank into matrices using compound indices,
mapping the pair of numbers $nn'$ to a single number $\nu$ (and similarly for $vv'$). Instead of (\ref{ME-tensor4})
we then solve the set of equations
\begin{equation}
\sum\limits_{\nu'} \mathcal{L}_{\nu\nu'} \rho_{\nu'} = 0
\end{equation}
together with the normalization condition
${\rm Tr}\{\rho\}\equiv\sum_n \rho_{nn}+\sum_v\rho_{vv}=1$
to find the stationary state.

\subsection{Observables of interest}

The general formula for the current (see, for example, \citet{hartle-2009}) through one level in ME theory reads
\begin{equation}
  I = \int\limits_0^{\infty} d\tau \text{Tr} \left\{
  [H_{ML}(-\tau), \rho \otimes \rho_l^0]
  \sum\limits_{k} V_{dk} (d^{\dagger} c_k - c_k^{\dagger} d)
  \right\},
\end{equation}
where only the contribution from the left lead is included in $H_{ML}$. In our case, the formula reduces to
\begin{eqnarray}
\label{current_mv}
 I=  \sum_{nv} \Gamma_{l}(\omega_{n v}) V^{l}_{n v}    \left\{
f_{l}(\omega_{n v}) \sum_{n'} V^{l}_{n' v} \rho_{n'n}  - \right. \nonumber \\
\left. -[1-f_{l}(\omega_{n v})] \sum_{v'} V^{l}_{n,v'}\rho_{vv'}  \right\}.
\end{eqnarray}
The mean value of the molecular Hamiltonian $H_M$ gives the average excitation energy of the bridge.
This quantity can be expressed as the sum of two contributions from the molecule occupied/unoccupied with
an additional electron, and in similar fashion as was the current
\begin{eqnarray}
 \langle H_M \rangle=\langle E_0\rangle+\langle E_1\rangle
                       =\sum_n E_n \rho_{nn} + \sum_v E_v \rho_{vv}.
\end{eqnarray}
The last important observable to be discussed is the angular momentum of the molecule. This is by construction
constrained along the rotational molecular axis z in our model. The corresponding operator reads
\begin{equation}
L_z = - i\frac{\partial}{\partial\varphi}.
\end{equation}
We notice, that the operator $L_z$ acts independently on the occupied and unoccupied bridge spaces, or, in other words,
the off-diagonal blocks $\rho_{01}$ and $\rho_{10}$ in the expansion (\ref{RDMdecomposition}) do not
contribute to the mean value. It allows us to write the mean value in the form
\begin{eqnarray}
\label{Lz_mean1}
\langle L_z \rangle = \sum\limits_{nn'}\langle n|L_z|n'\rangle\rho_{n'n} +
                      \sum\limits_{vv'}\langle v|L_z|v'\rangle\rho_{v'v} = \nonumber \\
                    = \sum_m m\rho^0_m + \sum_m m\rho^1_m,
\end{eqnarray}
where we have defined the populations $\rho^0_m$, $\rho^1_m$ of rotational eigenstates $L_z|m\rangle=m|m\rangle$
for unoccupied and occupied states of the molecule respectively, i.\ e.\
\begin{eqnarray}
  \rho^0_m &=& \sum\limits_{nn'}\langle m|n'\rangle \rho_{n'n}\langle n|m\rangle ,\\
  \rho^1_m &=& \sum\limits_{vv'}\langle m|v'\rangle \rho_{v'v}\langle v|m\rangle.
\end{eqnarray}
Notice the importance of the off-diagonal elements of the RDM in (\ref{Lz_mean1}). Since the diagonal
elements $\langle n|L_z|n\rangle$, $\langle v|L_z|v\rangle$ are zero we get $\langle L_z\rangle = 0$
if we neglect the off-diagonal elements $\rho_{nn'}$, $\rho_{vv'}$ for $n\neq n'$, $v\neq v'$.
While the value of the current is mainly determined by the diagonal elements $\rho_{nn}$, $\rho_{vv}$
the nonzero angular momentum is a consequence of the non-vanishing off-diagonal elements in
the energy representation, or equivalently a consequence of the asymmetry $\rho^i_m\neq\rho^i_{-m}$
in the angular momentum representation.

\section{Results and discussion}

In this section we discuss the results for the calculation of the current-voltage characteristics
and other properties of the junctions. We start with Model 1, where
the harmonic approximation holds for the small amplitude of vibrational motion and the levels are
near equidistant. In addition to the known behavior of such junctions with harmonic vibrations we can
expect some new effects due to the dependence of the molecule-lead coupling on the vibrational
coordinate. For any higher vibrational excitation of the junction molecule we would expect a  breaking of
the harmonic approximation.

\subsection{Current and excitation function}
\begin{figure}
\centerline{ \includegraphics[width=3.00in]{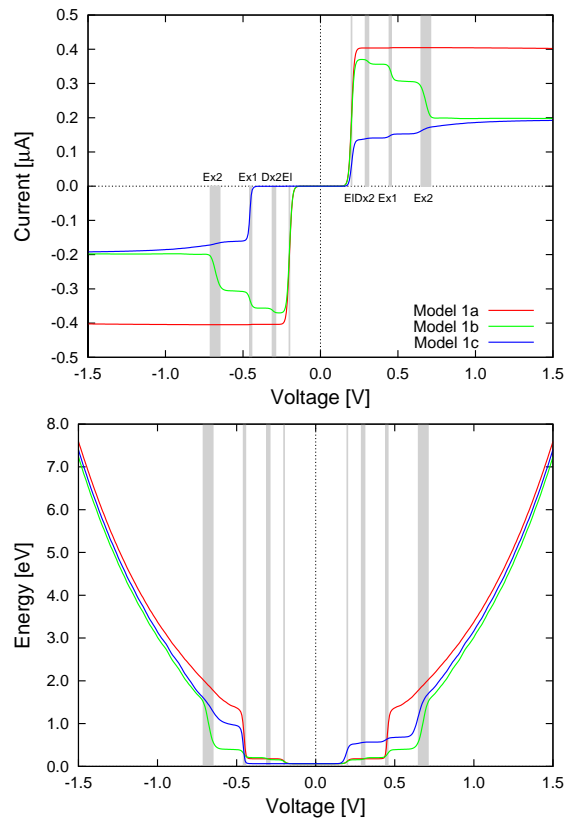} }
\caption{Current (top) and mean value of excitation energy $\langle H_M\rangle$ (bottom) as a function of the voltage, applied to the junction, for models 1a-c at temperature $T = 50 \,\text{K}$. The shaded bars show the positions of the steps derived from the energies of the molecular levels (see text).\label{m1_current_excitation}}
\end{figure}

The current-voltage characteristics and the vibrational excitation energy of the junction for Models 1a-c
at temperature $T=50\text{K}$ are shown in Figure \ref{m1_current_excitation}. Let us first focus on the current-voltage
curve. The red curve, for the model 1a, exhibits the behavior expected for the model with a very small
coupling between the vibrations and electronic motion. We observe a resonance step at the voltage
of $0.2$ V corresponding to twice the charging energy of the molecule $0.1$ eV. 

The models 1b (green line)
and 1c (blue line) differ by having a more complicated step structure and the different stationary value
of current that is finally reached. This second difference is easily understood: while the coupling to
the leads has the same maximum strength for all of the models
(they all reach the value $0.07$ eV), the angular dependence in models 1b, 1c makes it effectively smaller
(a quantitative argument as to why this difference is given by factor of two follows at the end of this section). 
The angular dependence is also a source of the negative differential conductance since the coupling in Model 1b reaches a maximum for angles near the equilibrium position of the vibrational
coordinate $\varphi$. The non-equilibrium vibrational distribution for larger voltages
therefore reduces the coupling.
The negative differential conductance effect disappears in Model 1c because the coupling 
to the right lead $V_R(\varphi)$ peaks at a different angle and is therefore effectively
increased by the increase in the vibrational excitation of the molecule.
Both Models
1b and 1c have the same average value of couplings $V_{L/R}(\varphi)$ and the asymptotic value
of the current for large voltages is therefore identical for both models.

\begin{table*}
  \caption{Inelastic one-electron attachment and detachment processes. The table gives the threshold electron energies of the inelastic processes and approximate size of the matrix elements responsible for the transitions due to electrons from the left and right leads. All values are in units of eV.\label{Series}}
  \begin{ruledtabular}
  \begin{tabular}{ccccc}
  Series &             Process                  & $\omega_{Nv}=E_v-E_n$ & $|\langle n|V_l|v\rangle|^2$ & $|\langle n|V_r|v\rangle|^2$ \\
  \colrule
   El    & $e_k+M_n^0\leftrightarrow M_{v=n}^1$ &         0.1           &             $0.2-1$          &        $\sim 10^{-4}$         \\
  \colrule
   Ex1   &$e_k+M_n^0\leftrightarrow M_{v=n+1}^1$&         0.21-0.23     &             $0.005-0.015$    &        $0.05-0.3$           \\
   Ex2   &$e_k+M_n^0\leftrightarrow M_{v=n+2}^1$&         0.31-0.36     &             $0.001-0.06$     &        $0.001-0.01$         \\
   Ex3   &$e_k+M_n^0\leftrightarrow M_{v=n+3}^1$&         0.42-0.48     &             $<0.005$         &        $<0.005$            \\
  \colrule
   Dx1   &$e_k+M_n^0\leftrightarrow M_{v=n-1}^1$&       -(0.01-0.03)    &             $0.005-0.015$    &        $0.05-0.3$         \\
   Dx2   &$e_k+M_n^0\leftrightarrow M_{v=n-2}^1$&       -(0.16-0.12)    &             $0.001-0.05$     &        $0.001-0.01$         \\
   Dx3   &$e_k+M_n^0\leftrightarrow M_{v=n-3}^1$&       -(0.23-0.28)    &             $<0.005$         &        $<0.005$            \\
  \end{tabular}
  \end{ruledtabular}
\end{table*}

To explain the details of the step-like behavior of the curves we start with
the mechanism of sequential tunneling through the bridge.
Thus electron conduction is understood as a sequence of charging (electron attachment)
\begin{equation}
e^- + M(n) \rightarrow M^-(v)
\label{TunnelingMechanism}
\end{equation}
and discharging (electron detachment)
\begin{equation}
M^-(v) \rightarrow M(n) + e^-
\label{TunnelingMechanism2}
\end{equation}
events on the bridge, where $e^-$ is the electron in the leads and $M(n)$ and $M^-(v)$ stands for the neutral molecule and the anion with the vibrational
states $|n\rangle$ and $|v\rangle$ respectively. In the first event (charging), the electron starts in one of the leads in state $|k\rangle$ and jumps into
the unoccupied bridge in the vibrational state $|n\rangle$ and turns it into an occupied bridge in the vibrational state $|v\rangle$. In the second event the electron starts in the occupied bridge in the state $|v\rangle$ and leaves the bridge in the state $|n\rangle$ jumping
into the leads to the state $|k\rangle$. The energy is conserved in both of the events
\begin{equation}
\varepsilon_k + E_n = E_v,
\end{equation}
where $\varepsilon_k$ is the energy of an electron in a state $|k\rangle$. 
The vibrational energies $E_n$ and $E_v$ of the occupied and unoccupied molecular bridge were defined in equation (\ref{e:h-s_equations}). 
Each difference $\omega_{nv} \equiv E_v-E_n$ defines a threshold energy for one possible in/out channel, which becomes available at the voltage $U={\pm}2\omega_{nv}$
(when the chemical potentials of the leads are equal $\mu_{\alpha}=\pm\frac{U}{2}=\pm\omega_{nv}$) and can
(but not necessary will) show itself as a step in Figure \ref{m1_current_excitation}. 
Since the structure of the energy levels is
the same for all the models 1a-c, we can expect the steps at the same voltage in all three current-voltage curves. 
The values of the threshold energies $\omega_{nv} \equiv E_v-E_n$ are shown
in Table \ref{Series}. 

The transitions are divided into several groups. The vibrational
state of the molecule is unchanged and $v=n$ in the first group denoted El. The corresponding threshold energy is $\omega_{nv}=0.1eV$
independent of the value of $n$ because the shape of the two potentials $V_0(\varphi)$
and $V_1(\varphi)$ is identical except for a vertical and horizontal shift. The excitation
groups Ex1-Ex3 correspond to $v=n+1$, $v=n+2$ and $v=n+3$. In the harmonic approximation the threshold
energies would also be independent of $n$. Table \ref{Series} shows the range of values of $\omega_{nv}$
for low states $n=0,1,...,10$ (the choice of this maximum value of $n$ is guided by the average excitation
energy shown in the lower part of the figure).
Lastly, the deexcitation groups Dx1-Dx3 are characterized by $v=n-1$,
$v=n-2$ and $v=n-3$. If we compare the predicted positions of the steps $U=\pm 2\omega_{nv}$ given
by the values from Table~\ref{Series} with the positions of the steps in Figure~\ref{m1_current_excitation},
we see that the steps correspond to voltages approximately 0.2, 0.3, 0.45 and 0.7 volts, i.\ e.\ to the series El, Dx2, Ex1 and Ex2.
It is remarkable that the vibrational energy of the bridge doesn't show any steps for a voltage $|U|>0.8\,\text{V}$
and grows parabolically for all models. The current-voltage characteristics also become smooth in this voltage region.
The level of excitation of the molecule is too high for the harmonic approximation to apply and
many different vibrational states are involved. The energy differences become irregular and 
it is not possible to sort them into groups, as in the case of the low, near harmonic vibrations above.
We would expect that quasi-classical theory can
be applied in this regime.

\begin{figure}
\centerline{ \includegraphics[width=3.00in]{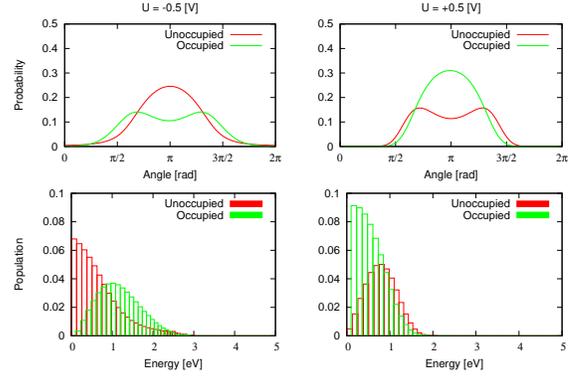} }
\caption{Angle distributions and populations calculated independently for the unoccupied and occupied bridge in Model 1c.\label{m1_voltage-asymmetry}}
\end{figure}
At equilibrium ($U=0$) only one lowest vibrational level $E_n^0=0.065\,\text{eV}$ is populated
(it is localized close to the minimum in the vibrational excitation curve
in Fig.~\ref{m1_current_excitation}). The bridge remains "closed" for current until the first step occurs at the voltage 0.2 V,
at which point the whole El set of channels then opens. If we consider the bridge originally in its ground state with energy
$E_n^0$ the first tunneling event
\begin{equation}
\varepsilon_k + E_{n=0} \xrightarrow{\mbox{\scriptsize ~~El~~}} E_{v=0}
\begin{array}{l}
\xrightarrow{\mbox{\scriptsize ~~El~~}} E_{n=0} + \varepsilon_k \\
\xrightarrow{\mbox{\scriptsize ~Dx1~}} E_{n=1} + \varepsilon_{k'}
\end{array}
\end{equation}
can leave the bridge in the excited state $E_{n=1}$ through the channel Dx1. When the next electron comes and the bridge is already in an excited state, tunneling can excite it even higher, but only by one quantum per tunneling event. Excitation to higher levels is limited by the competing process of de-excitation. This picture is the same for all three models. To appreciate the differences among
the models we must look at the Franck-Condon factors $|\langle n|V_l|v\rangle|^2$ and $|\langle n|V_r|v\rangle|^2$ responsible for the strength
of each $e_k+M_n^0\leftrightarrow M_{v}^1$ transition. 

It is $V_l=V_r=$constant for the Model 1a. Since the potentials $V_0$ and $V_1$
differ only slightly, it is $\langle n|v\rangle\simeq \delta_{nv}$ and the El channels are dominant with only a small contribution from
Ex1, Dx1 and virtually no contribution for higher channels. This explains why the red curves in the current-voltage and excitation graphs show
steps only at voltages corresponding to these channels. 

The values of $|\langle n|V_{\alpha}|v\rangle|^2$ for Models 1b and 1c 
are shown in Table~\ref{Series}. We should keep in mind that $V_l$=$V_r$ for Model 1b. The values of $|\langle n|V_{r}|v\rangle|^2$
shown in the last column are the ones for Model 1c. There is a pronounced difference between Models 1b and 1c in the first step at $U=0.2\,\text{V}$ and we can now understand why. While both charging and discharging of the bridge proceeds dominantly through the El channel for Model
1b (giving a step similar to that in Model 1a), the discharging to the right electrode through the channel El is strongly suppressed for Model 1c.
Thus for Model 1c discharging of the molecule to the right lead is possible mainly through the channel Dx1. This gives a
smaller value for the current, but a higher value of vibrational excitation for this model. Also, for higher voltages, the sizes of the steps in the
excitation curves follow the sizes of the Franck-Condon factors.

\begin{figure}
\centerline{ \includegraphics[width=3.00in]{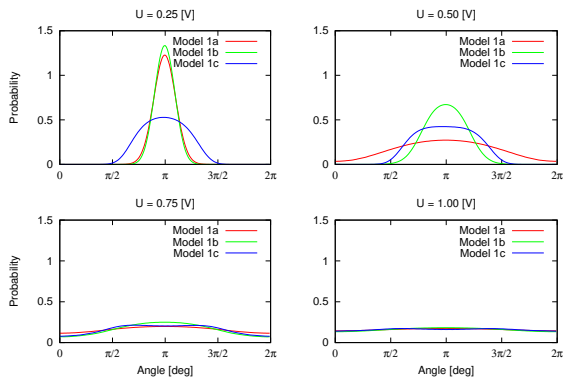}}
\caption{Angular probability distribution of the bridge $p(\varphi)$ at different voltages.\label{m1_angle-deloc}}
\end{figure}
One striking feature apparent from Figure~\ref{m1_current_excitation} is the asymmetry of the curves for Model 1c - a consequence of the asymmetry
$|\langle n|V_l|v\rangle|^2\neq|\langle n|V_r|v\rangle|^2$. For negative voltages the channel El is not available in Model 1c,
even for $U<-0.2\,\text{V}$, because charging has to proceed from the right electrode and the Franck-Condon factor $|\langle n|V_r|v\rangle|^2$
is suppressed by four orders of magnitude. Charging of the bridge only becomes possible with the availability of the Ex1 process. Both
current and vibrational excitation is thus only significant for negative voltages $U\lesssim -0.45\,\text{V}$.
Another way to look at this asymmetry is to inspect the angle distributions
\begin{eqnarray}
  p_0(\varphi) & \equiv & {\rm Tr} \{dd^{\dagger}\rho|\varphi\rangle\langle\varphi| \} \\
  p_1(\varphi) & \equiv & {\rm Tr} \{d^{\dagger}d\rho|\varphi\rangle\langle\varphi| \}
\end{eqnarray}
and the populations $\rho_{vv}$ of vibrational levels on the occupied and $\rho_{nn}$
on the unoccupied bridge, respectively. These are shown in Figure \ref{m1_voltage-asymmetry} for voltages $U={\pm}0.5\,\text{V}$.
We can make the observation that the angular distribution for the occupied molecule follows the shape of the angular dependence
of the coupling to the donor electrode (left/right for positive/negative voltage), and the distribution of angles for the unoccupied molecule 
follows the coupling to the acceptor electrode.
The same effect is responsible for the asymmetry found in the population distributions in the lower part of
Figure \ref{m1_voltage-asymmetry}. 
Similar distributions for Models 1a and 1b (not shown here) exhibit no asymmetry for the change $U\to -U$.
For these two models the vibrational distribution is also weakly correlated with the charging state of the molecule,
i.~e.\ the distributions $\rho_{nn}$ and $\rho_{vv}$ have almost identical shape as functions of energy $E_n$ and $E_v$ 
(i.~e.\ the red and green curves in Figure \ref{m1_voltage-asymmetry} are almost overlapping for Models 1a and 1b).
The small difference in the populations is only due to the small difference between the potentials $V_0$ and $V_1$.
The role of symmetries is further discussed in the next section.

The angular distributions $p(\varphi)={\rm Tr} \{\rho|\varphi\rangle\langle\varphi| \}$ at the voltages $U=0.25$, $0.5$, $0.75$ and $1.0$ volts  
are compared in Figure~\ref{m1_angle-deloc} for all the models 1a-c. There is a common trend for the angle
to become more and more delocalized while the voltage grows. The degree of delocalization follows from the excitation
curve in Figure \ref{m1_current_excitation} (lower graph). At $U=0.25\,\text{V}$, the distribution for Model 1c is the broadest, while
Model 1a wins out at higher voltages. The angle is
completely delocalized above the last step of the excitation in the IV curves. 
The current for Models 1b and 1c is asymptotically a factor of two smaller than the current for Model 1a. 
We have just seen that the angle distribution at the bridge is more or less homogeneous for large voltages. For this reason, the angle-dependent part of the coupling for Models 1b and 1c, which is equal to $(\cos(\varphi-\varphi_{\alpha}))^2$, reaches its mean value 0.5. For Model 1a, the angle-dependent part of the coupling is constant and equal to 1, i.e. twice as large.

\subsection{Angular momentum (motor effect)}

\begin{figure}
\centerline{ \includegraphics[width=3.00in]{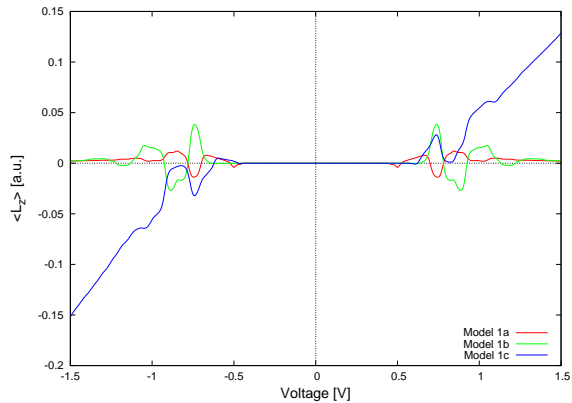} }
\caption{Mean value of the angular momentum of the molecule $\langle L_z \rangle$ as a function of voltage for Models 1a, 1b and 1c at the temperature $T = 50 \,\text{K}$\label{m1_momentum}}
\end{figure}

In Figure \ref{m1_momentum}, we plot the mean value of the angular momentum $\langle L_z \rangle$
against the voltage applied across the junction.
The calculated value of the angular momentum $\langle L_z \rangle$ is in general nonzero for all the models 1a-1c
and can reach values of the order of the reduced Planck constant $\hbar$ (atomic unit of angular momentum),
the value depending strongly on the voltage.
The model is thus an example of a molecular scale device
which can perform rotations controlled by the voltage applied to the device.
This effect is particularly pronounced in Model 1c, where the direction 
of rotation is inversed with the inversion of sign of the voltage.

The symmetries of the curves can be understood from the symmetries 
of each model. None of the models exhibit symmetry with respect to 
the inversion of the angle $\varphi\to -\varphi$, because this change 
inverts the horizontal shift of potential $V_1(\varphi)$ relative
to $V_0(\varphi)$. Breaking of this symmetry allows for nonzero values 
of angular momentum in all the models. But Models 1a and 1b are symmetric
with respect to the mutual interchange of the left and right lead. This is 
reflected in the symmetry of the angular momentum with respect to the voltage 
inversion $U\to -U$. Breaking of this symmetry makes Model 1c 
distinct. 

In all the models 1a-1c, the molecule does not rotate when the absolute value of the voltage is smaller than $0.4\,\text{V}$.
The onset of each curve follows a degree of vibrational excitation of the molecule (examine lower graph of Figure \ref{m1_current_excitation}).
This behavior reflects the fact that the molecule should be excited sufficiently high and overcome the rotational barrier at 2.5-2.6 eV
in order to perform rotational movement. A significant population of states above the barrier can only be expected 
when the mean value of the vibrational energy of the bridge molecule (lower graph, Figure \ref{m1_current_excitation}) is of the order of magnitude
of 1eV. 
The role of these "over-the-barrier" states was also checked by omitting all of the states below the barrier from the 
calculation. The linear growth of the angular momentum in Model 1c is hardly sensitive to this change. 
Another way to restate this discussion is to look again at the angle distribution functions in Figure \ref{m1_angle-deloc}.
The rotational motion of the molecule is indicated in the delocalization of the angle distribution through
the whole interval $\langle 0,2\pi\rangle$. At 0.5V, only Model 1a has this property, but at 0.75 V all three models can rotate.
These observations are in correspondence with Figure \ref{m1_momentum}.

\begin{figure}
\centerline{ \includegraphics[width=3.00in]{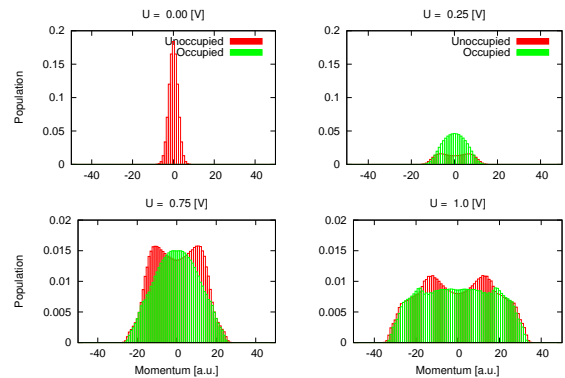} }
\caption{Populations $\rho^0_m$ and $\rho^1_m$ in the angular momentum basis for Model 1c at different voltages.\label{m1_populations-l}}
\end{figure}

The off-diagonal elements of the RDM are often neglected in the master equation approach, resulting in a set of equations
only for the population $\rho_{nn}$, $\rho_{vv}$ of the states. In fact, we used this line of thinking 
when discussing the current-voltage and excitation curves in the previous section.
If we want to capture the motor effect we have to consider the off-diagonal elements - at least for the near degenerate
states close to or above the rotational barrier. 
This matter we discussed already in the Theory section below Formula (\ref{Lz_mean1}), 
where we also saw that the coefficients $\rho^0_m$ and $\rho^1_m$
give a better insight into the calculation of $\langle L_z\rangle$.
These coefficients are shown in Figure \ref{m1_populations-l} for four voltages.
The nonzero mean value of the angular momentum is a consequence of the asymmetry of these distributions with respect to $m=0$.
For example, the last two graphs at voltages $U = 0.75\,\text{V}$ and $U = 1.0\,\text{V}$ have a small asymmetry (about 1 per cent,
not possible to see in the figure). This asymmetry will be much more pronounced in Model 2.

\begin{figure}
\centerline{ \includegraphics[width=3.00in]{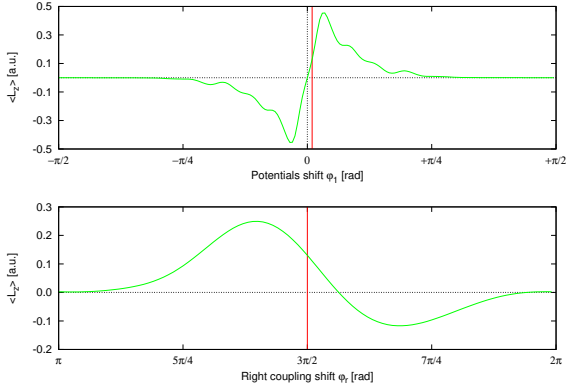} }
\caption{Dependence of the mean value of angular momentum $L_z$ at voltage $U=1.5\,\text{V}$ on the model parameters of potential shift $\varphi_1$ (top) and coupling asymmetry $\varphi_r$(bottom) in Model 1c.\label{m1_symmetry-phi}}
\end{figure}

At the beginning of this section, we discussed the role of symmetry in the models with respect
to the transformation $\varphi\to -\varphi$ and $U\to -U$. The parameters responsible for the asymmetry
of the models with respect to these two transformations are the charged bridge potential shift $\varphi_1$ 
and the right coupling shift $\varphi_r$. 
In Figure \ref{m1_symmetry-phi}, we show $\langle L_z \rangle$ as a function of $\varphi_1$  and $\varphi_r$
for Model 1c with the voltage fixed at $U=1.5\,\text{V}$;
the red line in the figure marks the values actually used in Model 1c.
The angular momentum first grows with $\varphi_1$ but higher shifts suppress the effect again. This effect is similar to the Frank-Condon blockade for the current observed in \citet{koch-2005}. Selecting the optimum value
for this parameter can enhance the angular momentum by a factor of three.
The dependence of $\langle L_z \rangle$ on $\varphi_r$ is shown in the lower graph of Figure \ref{m1_symmetry-phi}.
It is possible to maximize the "motor effect" by optimizing $\varphi_r$, gaining another factor of 2.
Coupling becomes symmetrical when $\varphi_r = \pi$ or $\varphi_r = 2\pi$ as in Model 1b, where
$\langle L_z \rangle$ vanishes.

\subsection{Results for the model 2}

We now discuss the more realistic model~2. Here the difference between the shapes of the potentials for the charged and uncharged molecule
is large, leading to strong coupling and the impossibility to sort the vibrational transitions into more or
less sharp lines. Furthermore, we took a more realistic value for the moment of inertia (corresponding to benzene), leading to a very dense
vibrational spectrum. The calculation is therefore much more numerically demanding and we had to 
cover a smaller voltage range.

\begin{figure}
\centerline{ \includegraphics[width=3.00in]{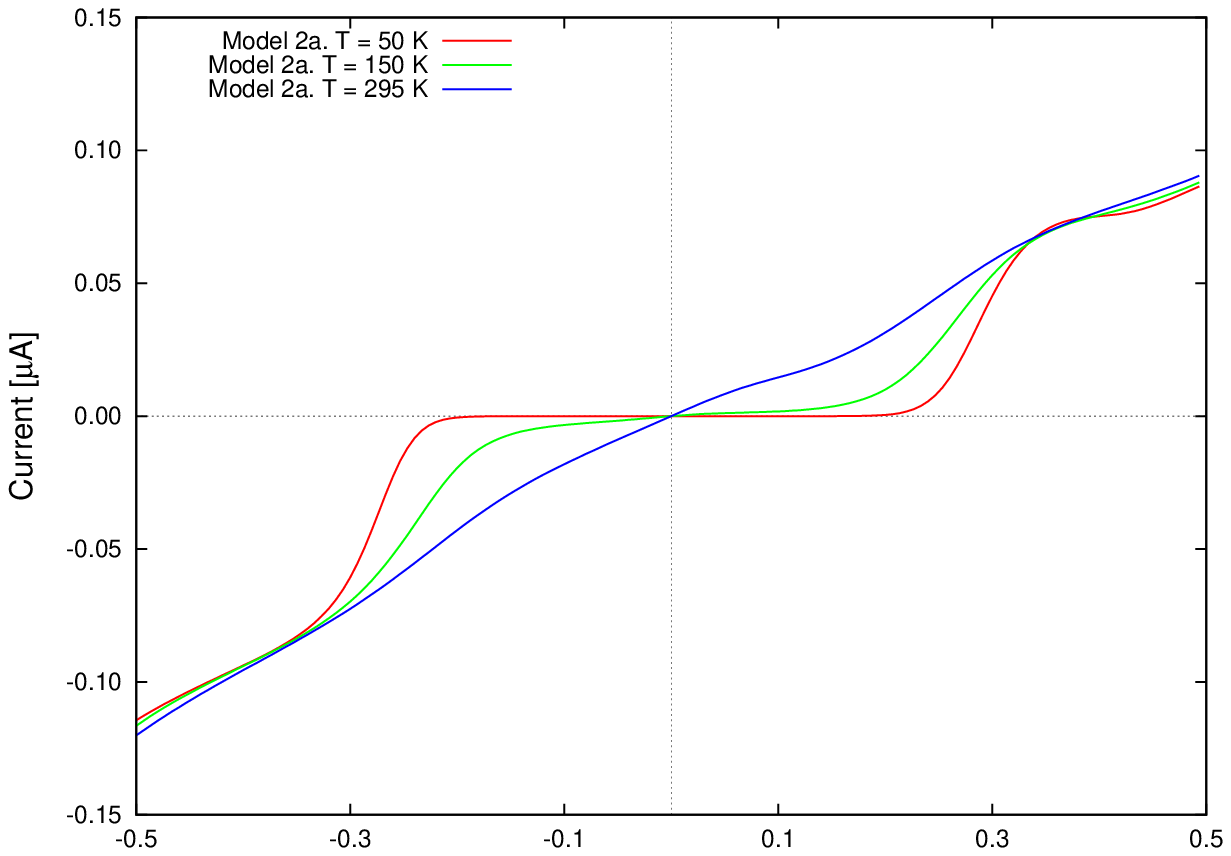} }
\centerline{ \includegraphics[width=3.00in]{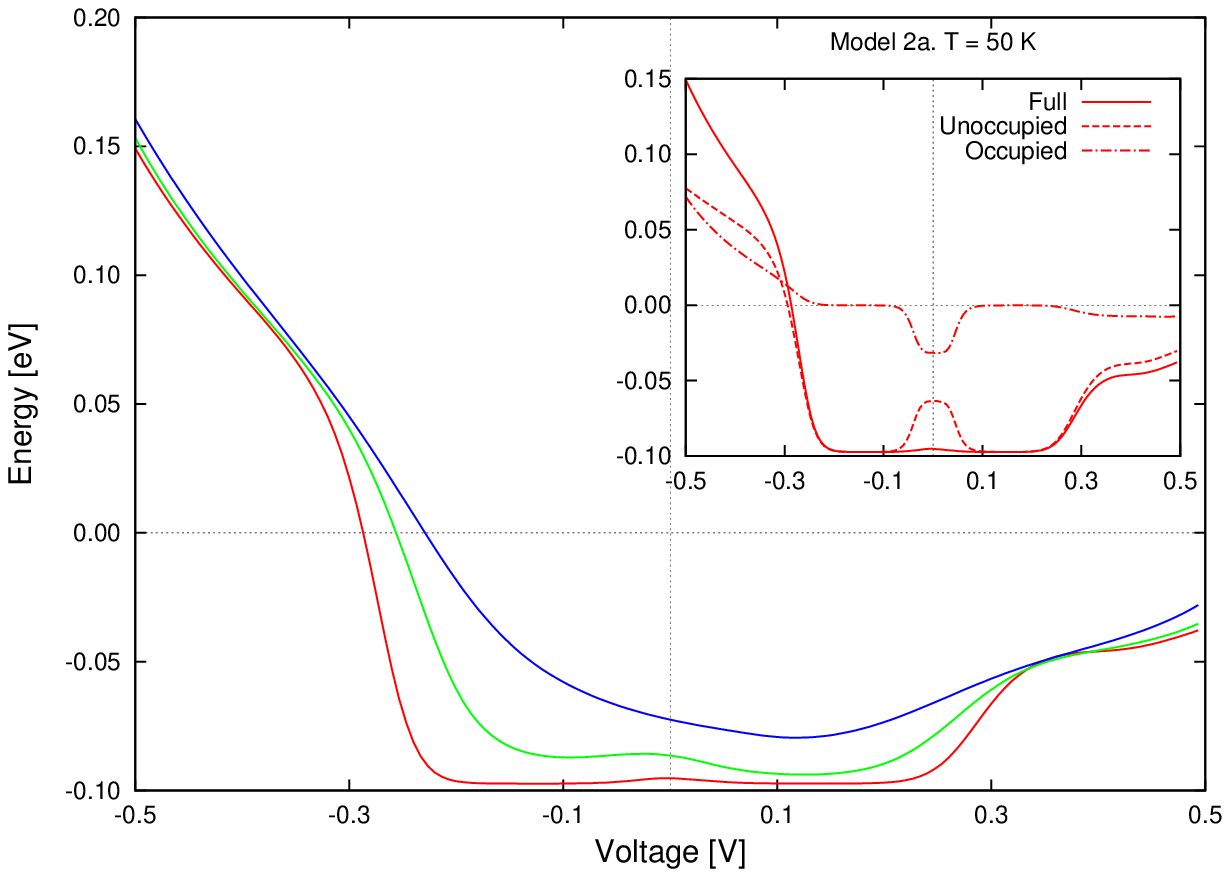} }
\caption{The current and the excitation function for Model 2a are plotted as a function of the bias voltage.
The results for three different temperatures (including room temperature) are shown. The inset in the bottom part shows
the contribution to vibrational energy from the unoccupied and occupied bridge for the temperature 50 K. \label{m2_current}}
\end{figure}

The current and the excitation function of the molecule for three different temperatures are plotted in Figure \ref{m2_current}
as a function of voltage. As we have already mentioned, the density of the possible tunneling channels in Model 2 is much higher 
than in Model 1. We therefore expect no distinct steps
in the voltage dependencies of observables for this model.
Another difference is that the ground state energies of the unoccupied
and occupied bridge almost coincide in Model 2. 
This means that the bridge is open for electrons with zero energy and tunneling events
can happen at zero voltage.
Despite the large amount of possible channels, we observe a zero current plateau in Fig.\ \ref{m2_current}.
This is again a consequence of the small values of the Frank-Condon overlaps for the low-lying vibrational states in both potentials.
The current plateau disappears for the calculation at higher temperatures, where higher states, not subjected
to the Frank-Condon blockade, are already excited at zero voltage.

Temperature effects become much more important for Model 2 than it was for Model 1. This is connected with the fact that even at the lowest temperatures considered here (50K), $kT$ is comparable with distances between energy levels of the molecule.
It is also interesting to note, that the vibrational excitation curve in
the lower graph of figure \ref{m2_current} does not have the minimum at zero voltage, but at
approximately $U=\pm 0.15\,\text{V}$. This implies the presence of a cooling effect of the current, which has already been observed experimentally \cite{ioffe-2008} and also been discussed theoretically \cite{zippilli-2009, galperin-2009, hartle-2009}.

\begin{figure}
\centerline{ \includegraphics[width=3.00in]{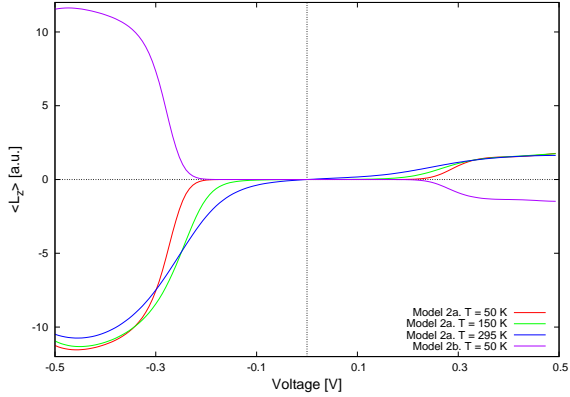} }
\caption{Mean value of the angular momentum of the molecule $\langle L_z \rangle$ as a function of voltage for Models 2a and 2b. The temperature dependence is also shown for Model 2a.\label{m2_momentum}}
\end{figure}

Both the excitation function and
the angular momentum voltage dependence (see Figure \ref{m2_momentum}) are strongly asymmetric, which is not surprising, for the strongly asymmetric
coupling.  It is also obvious from Figure~\ref{m2_potentials} that the left lead coupling strength is minimal
in the area where the unoccupied bridge wave functions are localized, which makes the overall coupling of the left lead smaller
compared to the right lead.

The angular momentum reaches much higher values
(up to  $12\hbar$) inside the considered voltage interval, which could partially be attributed to the high moment
of inertia. Temperature has some influence on the shape of the angular momentum curve but its maximum
value is rather insensitive to temperature. From this we can conclude
that the motor effect can be observed at room temperatures as well as at cryogenic temperatures.

\begin{figure}
\centerline{ \includegraphics[width=3.00in]{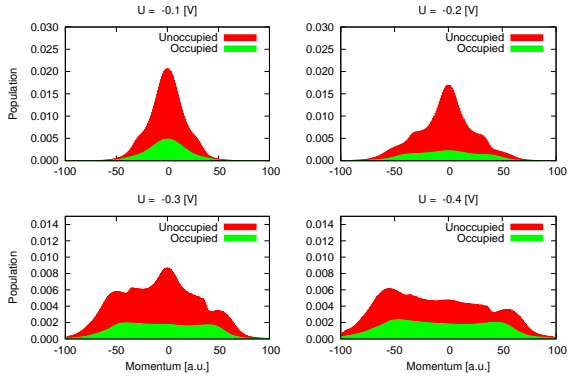} }
\caption{Populations distributions $\rho^0_m$ and $\rho^0_v$ of the RDM in momentum basis are plotted for different bias voltages. It is clearly seen how distributions become wider and lose their symmetry while absolute values of the voltage grow. Graphs are plotted for room temperature (295 K).\label{m2_populations-l}}
\end{figure}

As we discussed before, the nonzero mean values of the angular momentum are connected with asymmetries in the population distributions
of the eigenstates of $L_z$, which are plotted in Figure \ref{m2_populations-l}.
Large values of $\langle L_z \rangle$ are accompanied by large asymmetries, which are clearly seen
for $U = -0.3\,\text{V}$ and $U = -0.4\,\text{V}$.


\section{Conclusions and future prospects}

In this paper we have discussed the models of a molecule coupled to two conducting leads with
coupling depending on the vibrational coordinate. It was shown how the steps in the current-voltage
characteristics can be analyzed in the case of anharmonic vibrations of the molecule.
We have also calculated the population of molecular vibrational states for different
voltages. In the case of the model which was asymmetric with respect to an exchange
of the left and the right leads, we had to distinguish between the two charging states of the molecule,
which have different populations, i.\ e.\ the population of vibrational states was strongly
correlated with the charging state of the molecule.

In addition, we have studied the "motor effect", i.\ e.\ the response of the angular momentum
of the molecule to the voltage applied across the junction. We have demonstrated that
the mean value of the angular momentum strongly depends on the voltage. 
For the asymmetrically coupled molecule, the direction of the molecular rotations can be controlled 
by the polarity of the voltage. Significant values of angular momentum were reached when 
the vibrational levels above the potential energy barrier against full rotation were populated.

It will be interesting to see the effect of the higher order terms 
(see, for example, \citet{esposito-2010}),
as well as friction, to the dynamics of the molecular rotor in future work.
We believe that the "motor effect" must survive these additional corrections
since it is quite stable with respect to changes to the model and it is 
a consequence of the breaking of symmetry of the junction with respect to the inversion 
of the angular coordinate and the interchange of the left and right electrode
(chirality of the junction). 

\begin{acknowledgments}
This work was supported by the Charles University Grant Agency (GAUK 116-10/143107) and Grant Agency of Czech Republic (GACR 208/10/1281).
We thank Michael Thoss, Wolfgang Domcke, Maarten Wegevijs, Tom\'{a}\v{s} Novotn\'{y} and Rainer H\"artle for fruitful discussions and Steve Ridgill for text corrections.
\end{acknowledgments}

\bibliography{motor}

\end{document}